\documentclass[10pt,conference]{IEEEtran}
\pdfoutput=1 
\usepackage{amssymb}
\usepackage{amsmath}
\usepackage{amsopn}
\usepackage{graphicx,color}                   
\usepackage{rotating}
\usepackage{bm}
\usepackage{cite}          

\allowdisplaybreaks

\newcommand{\comment}[1]{}  
\definecolor{old}{gray}{0.6}

\renewcommand{\vec}[1]{\mathbf{#1}}
\newcommand{\mat}[1]{\mathbf{#1}}

\newcommand{\dotequal}{\stackrel{.}{=}}
\newcommand{\dotleq}{\stackrel{.}{\leq}}
\newcommand{\dotgeq}{\stackrel{.}{\geq}}

\renewcommand{\P}{\text{P}}

\newcommand{\Rnum}{{\mathbb R}}
\newcommand{\Cnum}{{\mathbb C}}
\newcommand{\Znum}{{\mathbb Z}}

\newcommand{\CZnum}{{\mathbb C}{\mathbb Z}}

\newcommand{\ILSP}{\text{CLP}}

\newcommand{\rcover}{\mu}
\newcommand{\rtwo}{\rho}

\renewcommand{\L}{L}
\newcommand{\lattice}{{\cal L}}

\newcommand{\Qb}{\mat{Q}}
\newcommand{\Qbr}{\widetilde{\mat{Q}}}
\newcommand{\Rbr}{\widetilde{\mat{R}}}
\newcommand{\Hbr}{\widetilde{\mat{H}}}

\newcommand{\RbrR}{\mat{R}'}
\newcommand{\QbrR}{\mat{Q}'}

\newcommand{\ybr}{\widetilde{\vec{y}}}
\newcommand{\Tb}{{\mat{T}}}

\newcommand{\CN}{{\cal C}{\cal N}}

\newcommand{\NN}{S}
\newcommand{\NNtotal}{S}

\newcommand{\cexp}{\xi}
\newcommand{\cexptotal}{\xi}

\newcommand{\layerk}{k}

\newcommand{\tdim}{M}
\newcommand{\rdim}{N}

\newcommand{\Hb}{\mat{H}}

\newcommand{\db}{\vec{d}}

\newcommand{\rb}{\vec{r}}
\newcommand{\wb}{\vec{w}}

\renewcommand{\d}{d}
\newcommand{\wvar}{{\sigma^2}}

\newcommand{\Rb}{\mat{R}}

\newcommand{\yb}{\vec{y}}
\newcommand{\y}{y}

\title{Tail Behavior of Sphere-Decoding Complexity\\ in Random Lattices}

\author{ 
\IEEEauthorblockN{D.~Seethaler\IEEEauthorrefmark{1},  J.\ Jald\'en$^{\circ}$, C.~Studer\IEEEauthorrefmark{1}, and H.~B\"olcskei\IEEEauthorrefmark{1}} 
\vspace{2mm}
\IEEEauthorblockA{\IEEEauthorrefmark{1}
Communication Technology Laboratory\\[-0.5mm] ETH Zurich, 8092 Zurich, Switzerland, email:\{seethal,studerc,boelcskei\}@nari.ee.ethz.ch } 
\vspace{0.5mm}
\IEEEauthorblockA{$^{\circ}$
Institute of Communications and Radio-Frequency Engineering\\[-0.5mm]
Vienna University of Technology, 1040 Vienna, Austria, email: jjalden@nt.tuwien.ac.at} \\[-12mm]
 \thanks{This work was supported in part by the STREP project No. IST-026905 (MASCOT) within the Sixth Framework Programme of the European Commission.}
} 

\comment{\author{ 
\IEEEauthorblockN{D.~Seethaler\IEEEauthorrefmark{1},  J.\ Jald\'en$^{\circ}$, C.~Studer\IEEEauthorrefmark{1}, and H.~B\"olcskei\IEEEauthorrefmark{1}} 
\thanks{This work was supported in part by the STREP project No. IST-026905 (MASCOT) within the Sixth Framework Programme of the European Commission.}
}} 

\IEEEoverridecommandlockouts 

\begin{document}

\maketitle

\renewcommand{\baselinestretch}{0.975}\small\normalsize

\begin{abstract}
We analyze the (computational) complexity distribution of sphere-decoding (SD) for random infinite lattices. In particular, we show that 
under fairly general assumptions on the statistics of the lattice basis matrix, the tail behavior of the SD complexity distribution is solely determined by the inverse volume of a fundamental region of the underlying lattice. Particularizing this result to $\rdim \times \tdim$, $\rdim \geq \tdim$,  i.i.d.\ Gaussian lattice basis matrices, we find that the corresponding complexity distribution is 
of Pareto-type with tail exponent given by $\rdim-\tdim+1$. We furthermore show that this tail exponent is not improved by lattice-reduction, which includes layer-sorting as a special case.
\end{abstract}

\section{Introduction}
The problem of finding the closest lattice point in an infinite lattice 
is commonly referred to as the {\em closest lattice point} ($\ILSP$) problem (see, e.g., \cite{agrell_it02}). The sphere-decoding (SD) algorithm \cite{fincke_phost85,mow92,viterbo93,agrell_it02,burg05_vlsi,studer08} 
is a promising approach for solving the $\ILSP$ problem. The (computational) complexity of SD, as measured in terms of the number of  searched lattice points, depends strongly 
on the lattice basis matrix and is, in general, very difficult to characterize analytically. However, if the basis matrix is assumed {\em random}, the complexity of SD is random as well and one can resort to a characterization of the {\em complexity distribution} of SD. Previous work along these lines focused on the mean and the variance of SD complexity for i.i.d.\ Gaussian basis matrices 
\cite{hass_sp03_part_i,hass_sp03_part_ii,jalden_tsp05}. However, a characterization of the tails of the SD complexity distribution is, for example, important for using SD 
under 
run-time constraints (see, e.g., \cite{studer08}). In this paper, we make a first attempt in this direction by analyzing the {\em tail behavior} (TB) of the 
SD complexity distribution.
Our main contributions can be summarized as follows: 
\begin{itemize}
\item Under fairly general assumptions on the statistics of the lattice basis matrix, we prove that the TB of the SD complexity distribution 
	 is solely determined by the TB of the inverse volume of a fundamental region of the underlying lattice.
\item Specializing this result to the case of $\rdim \times \tdim$,  $\rdim \geq \tdim$, i.i.d.\ circularly symmetric complex Gaussian basis matrices, we find that the complexity distribution of SD is 
	of Pareto-type with tail exponent given by $\rdim-\tdim+1$. We furthermore show that the tail exponent is not improved, i.e., increased, if lattice-reduction (see, e.g., \cite{agrell_it02}), which includes layer-sorting as a special case, is performed.
\end{itemize}

\subsubsection*{Notation}\label{sec.notation}
We write $A_{i,j}$ for the entry in the $i$th row and $j$th column of the matrix $\vec{A}$ and $x_i$ for the $i$th entry of the vector $\vec{x}$. 
Slightly abusing common terminology, we call an $\rdim \times \tdim$, $\rdim \geq \tdim$, matrix $\vec{A}$ unitary if it satisfies $\vec{A}^{\!H}  \vec{A} = \vec{I}$, where $^H$ denotes conjugate transposition
 and $\vec{I}$ is the identity matrix. 
The Euclidean- and the Frobenius norm are denoted by $\|\cdot\|$ and $\|\cdot\|_{\text{F}}$, respectively, and $|{\cal X}|$ refers to the cardinality of the set ${\cal X}$. The 
ceil-function is denoted by $\lceil \cdot \rceil$. 
Furthermore, $\CZnum$ stands for the set of Gaussian integers, i.e., $\CZnum = \Znum + \sqrt{-1}\,\Znum$. The lattice generated by the full-rank $\rdim\times \tdim$ ($\rdim \geq \tdim$) 
basis matrix $\vec{A}$ is defined as $\lattice(\vec{A}) \,=\, \big\{ \vec{A} \db:\, \db \in (\CZnum)^{\tdim}\big\}$. 
For $\rdim = \tdim$, the corresponding covering radius is given by \cite{conway88}
\begin{equation}\label{eqn.covering.radius}
\mu(\vec{A}) =  \underset{\vec{x} \in \Cnum^\tdim} { \mbox{max}}\, \underset{\db \in (\CZnum)^\tdim} { \mbox{min}}\,
\|\vec{x} -\vec{A} \db\|. 
\end{equation}
A circularly symmetric complex Gaussian random variable (RV) with variance $\sigma_{x}^2$ is denoted as $x \sim  \CN(0,\sigma_{x}^2)$.  
The natural logarithm is referred to as $\text{log}(\cdot)$. 
 We write $g(x) \dotequal f(x)$, $x\rightarrow x_{0}$, for 
$\text{lim}_{x \rightarrow x_{0}} \,\text{log}\, g(x)/\text{log}(x) = \text{lim}_{x \rightarrow x_{0}} \,\text{log}\, f(x)/\text{log}(x)$, assuming that the corresponding limits exist. 
\sloppy The symbols $\dotleq$ and $\dotgeq$ are defined analogously. Finally, non-polynomial behavior of $g(x)$ is captured by 
$\text{lim}_{x \rightarrow x_{0}} \text{log}\,g(x)/\text{log}(x) = \pm\infty$ or $\text{lim}_{x \rightarrow x_{0}} \text{log}\,g(x)/\text{log}(x) = 0$, for which we write 
$g(x) \dotequal x^{\pm \infty}$, $x\rightarrow x_{0}$, and $g(x) \dotequal x^{0}$, $x\rightarrow x_{0}$, respectively. 

\subsection{Sphere-Decoding}\label{sec.ILSP}

The $\ILSP$ problem refers to computing 
\begin{equation}\label{eqn.ILSP}
\widehat{\db} \,=\, \underset{\db \in (\CZnum)^\tdim} { \mbox{arg\, min}}\,
\|\rb-\Hb\db\|^2
\end{equation}
for a given vector $\rb\in \Cnum^{\rdim}$ and a given full-rank matrix  $\Hb \in \Cnum^{\rdim\times \tdim}$, $\rdim \geq \tdim$. 
In words, solving \eqref{eqn.ILSP} amounts to finding the point in the 
lattice $\lattice(\Hb)$  that is closest (in Euclidean distance) to $\rb$. In communications, \eqref{eqn.ILSP} is known as the  
maximum-likelihood (ML) detection problem for detecting $\db' \in  (\CZnum)^\tdim$  based on the linear model $\rb \,=\, \Hb\db' + \wb$ with $\Hb$ known at the receiver and $\wb$ being i.i.d.\ circularly symmetric complex Gaussian noise.

A prominent approach for solving \eqref{eqn.ILSP} is the SD algorithm  \cite{fincke_phost85,mow92,viterbo93,agrell_it02,burg05_vlsi,studer08}. In the following, we consider Fincke-Pohst SD \cite{fincke_phost85}  %
without radius reduction  (see, e.g., \cite{hass_sp03_part_i}). The algorithm starts by computing the (unique) QR-decomposition (QRD) $\Hb = \Qb\Rb$, where  
 $\Rb$ denotes an $\tdim \times \tdim$ upper triangular matrix with positive real-valued elements on its main diagonal and  $\Qb$ of size $\rdim\times\tdim$ is unitary. Then, \eqref{eqn.ILSP} can equivalently be written as
\begin{equation}\label{eqn.ILSP.QR}
\widehat{\db} \,=\, \underset{\db \in (\CZnum)^\tdim} { \mbox{arg\, min}}\,\|\yb -   \Rb \db \|^2 
\end{equation}
where $\yb = \Qb^H \rb$. Next, \eqref{eqn.ILSP.QR} is solved subject to a  {\em sphere constraint} (SC), which amounts to considering only those $\db \in (\CZnum)^\tdim$ that lie within a hypersphere of radius $\rtwo$ around $\yb$, i.e., all $\db$ that satisfy 
\begin{equation}\label{eqn.SC.qr}
    \|\yb -   \Rb \db \|^2 \leq \rtwo^2. 
\end{equation}
Here, the sphere radius $\rtwo$ has to be chosen sufficiently large for the search sphere to contain at least one lattice point $\Rb \db$. Note, however, that if $\rtwo$ is chosen too large, too many points will satisfy the SC and the complexity of SD will be high.  As detailed next, 
imposing a SC enables an efficient recursive solution of the triangularized $\ILSP$ problem \eqref{eqn.ILSP.QR}. 

Consider the length-$\layerk$ subvectors $\db_\layerk \in (\CZnum)^\layerk$ of $\db$ defined as 
$\db_\layerk = (\d_{\tdim-\layerk+1} \cdots\, \d_{\tdim})^T$, 
$\layerk = 1, \dots, \tdim$, 
where $\layerk$ is a {\em layer} index. 
The metric $\|\yb -   \Rb \db \|^2 =  \|\yb_{\tdim} -   \Rb_{\tdim} \db_{\tdim} \|^2$ can be computed recursively 
according to 
\begin{equation}\label{eqn.metric.update}
\|\yb_{\layerk} -   \Rb_{\layerk} \db_{\layerk}  \|^2 = \|\yb_{\layerk-1} -   \Rb_{\layerk-1} \db_{\layerk-1} \|^2 
+ |\Delta_{\layerk}(\db_{\layerk})|^2
\end{equation}
 where 
$\Delta_{\layerk}(\db_{\layerk}) = y_{\tdim-\layerk+1} - \sum_{i=\tdim-\layerk+1}^\tdim R_{\tdim-\layerk+1,i} \,d_i$, 
 $\Rb_\layerk$ denotes the  $\layerk \times \layerk$ bottom right (upper triangular) submatrix of $\Rb$ associated with $\db_\layerk$, and 
$\yb_\layerk = (\y_{\tdim-\layerk+1} \cdots\, \y_{\tdim})^T$.   Thus, with \eqref{eqn.metric.update}, a necessary condition for  $\db$ to satisfy the SC is that any associated $\db_\layerk$ satisfies the {\em partial SC}   
\begin{equation}\label{eqn.PSC}
\|\yb_{\layerk} -   \Rb_{\layerk} \db_{\layerk}  \|^2  \leq \rtwo^2. 
\end{equation} This formulation now enables finding all integer vectors $\db$ that satisfy \eqref{eqn.SC.qr} 
in an efficient (recursive) manner as detailed, e.g., in  \cite{hass_sp03_part_i,burg05_vlsi}. 

\comment{ Starting with layer $\layerk = 1$, \eqref{eqn.PSC} yields 
\begin{equation}\label{eqn.PSC.k1}
 	|y_{\tdim} - R_{\tdim\!,\tdim} d_{\tdim}|^{2} \leq  \rtwo^2
\end{equation}
stating that the $\tdim$th component $d_{\tdim}$ of all $\db \in (\CZnum)^\tdim$ satisfying \eqref{eqn.SC.qr} is 
located inside a circle of radius $\rtwo/{R_{\tdim\!,\tdim}}$ with center point $y_{\tdim}/{R_{\tdim\!,\tdim}}$. For every  $d_{\tdim} \in \CZnum$ satisfying 
\eqref{eqn.PSC.k1}, in the next step, one finds all  $d_{\tdim-1} \in \CZnum$ such that \eqref{eqn.PSC} is satisfied for $\layerk = 2$. 
This procedure is repeated until $\layerk = \tdim$.  Among the lattice points delivered by the algorithm, the one 
with minimum $\|\yb - \Rb  \db  \|^2$ constitutes the solution of \eqref{eqn.ILSP.QR}. }

\section{Complexity Distribution of SD}
We define the computational complexity of SD as the number of lattice points searched by the algorithm, i.e., the number of vectors $\db_{\layerk} \in (\CZnum)^{\layerk}$, $\layerk = 1,\dots,\tdim$, that satisfy the 
partial SCs in 
\eqref{eqn.PSC} 
(cf.\ \cite{banih_98,hass_sp03_part_i}). 
Specifically, we define the $\layerk$th {\em layer complexity} of SD as 
\begin{equation}\label{eqn.def.Sk}
	\NN_{\layerk} = \left|\left\{\db_{\layerk} \in (\CZnum)^{\layerk}:\, \|\yb_{\layerk} -   \Rb_{\layerk} \db_{\layerk}  \|^2  \leq \rtwo^2 \right\}\right|
\end{equation}
with the  {\em total complexity} 
$\NNtotal  =  \sum_{\layerk=1}^{\tdim} \NN_{\layerk}$. 
It was shown in \cite{burg05_vlsi} -- for  the finite lattice case -- that $\NNtotal$ is proportional to the run-time complexity of a corresponding VLSI implementation. 

\subsection{Complexity Distribution and Tail Exponents}
The quantities $\NN_{\layerk}$,  $\layerk = 1,\dots,\tdim$, and $\NNtotal$, defined above, are functions of $\Hb$, $\rb$, and $\rtwo$. 
In the following, we let $\Hb$ and $\rb$ be random (potentially statistically dependent) and consider a fixed $\rho$ that does not depend on the realizations of $\Hb$ and $\rb$. 
For example, in the case of ML detection of the transmitted data vector $\db'$ in 
MIMO wireless systems, $ \rb \,=\, \Hb\db' + \wb$, where the entries of $\Hb$ (the channel matrix) and $\wb$ (the noise vector) are typically assumed 
i.i.d.\  circularly symmetric complex Gaussian. In this setting, a reasonable fixed choice 
of  $\rtwo$ can be based on the noise statistics such that the probability of finding the transmitted data vector inside the search hypersphere is sufficiently high (see, e.g., 
\cite{hass_sp03_part_i}). Note, however, that this results in a nonzero probability of failing to find an integer vector 
inside the search hypersphere, which, in the absence of restarting the search with a larger sphere radius (see, e.g., \cite{hass_sp03_part_i}), would entail an error floor. 

Since both $\Hb$ and $\rb$ are random, 
$\NN_{\layerk}$ and $\NNtotal$ are random as well and can be characterized through their  respective {\em distribution functions}  $\P[ \NN_{\layerk} \geq \L ]$ and 
$\P[ \NNtotal \geq \L ]$. While these distributions seem hard to come by analytically, it turns out that the corresponding {\em tail exponents} $\cexp_{\layerk}$, $\layerk = 1,\dots, \tdim$, and $\cexp$, defined by 
\begin{equation*}
 \P[ \NN_{\layerk}  \geq \L ]  \dotequal \L^{-\cexp_{\layerk}}, \quad  \L\rightarrow \infty 
 \end{equation*}
 and 
\begin{equation*}
 \P[ \NN  \geq \L ]  \dotequal \L^{-\cexp}, \quad \L\rightarrow \infty
\end{equation*}
are amenable to an analytical characterization. We note that 
$\NNtotal  =  \sum_{\layerk=1}^{\tdim} \NN_{\layerk}$ implies 
\begin{equation}\label{eqn.cexp.total.cexpk}
	\cexptotal = \text{min}\{\cexp_{1},\cexp_{2}, \dots, \cexp_{\tdim} \}.
\end{equation}
The tail exponents characterize the TB of the corresponding complexity distributions in terms of polynomial decay rates in $\L$ for $\L \rightarrow \infty$. 
We note that for finite (non-zero) tail exponents, the corresponding complexity distributions are of Pareto-type meaning that they decay polynomially in $\L$. In particular, 
if the complexity distribution $\P[ \NN  \geq \L ]$ has tail exponent $\cexp$, one can state that 
 $\L^{-(\cexp+\delta)} \leq \,\P[ \NN  \geq \L ]\, \leq \L^{-(\cexp-\delta)}$
  for any $\delta> 0$ and sufficiently large $\L$.  Furthermore, if $\P[ \NN^{(1)}  \geq \L ]$ and $\P[ \NN^{(2)}  \geq \L ]$ have tail exponents 
  $\cexp^{(1)}$ and $\cexp^{(2)}$ with  $\cexp^{(1)} > \cexp^{(2)}$,  we can conclude that $\P[ \NN^{(1)}  \geq \L ] < \P[ \NN^{(2)}  \geq \L ]$ for 
  sufficiently large $\L$. Hence, larger tail exponents are desirable since this implies that   
the probability of the complexity being atypically large is smaller.  This, for example, is advantageous in the context of MIMO detection under run-time constraints (i.e., under limits on the number of lattice points that can be searched). We emphasize, however, that the complexity tail exponents, as defined above, do not capture multiplicative constants and do not characterize the small $\L$ behavior of the corresponding 
complexity distributions. 
\comment{This is 
conceptional similar to the diversity analysis of the error probability of data detection in wireless communications 
(see, e.g., \cite{zheng02}), where, for example, no potential 
coding gains are captured.}

\subsection{Main Result}

The complexity of SD can often be reduced by employing preprocessing techniques such as lattice-reduction (LR) or layer-sorting (LS) (see, e.g.,  \cite{agrell_it02}). In the remainder of this paper, we account for preprocessing by assuming that $\yb$ is a general function of $\rb$ and $\Hb$ and $\Rb$ is a general function of $\Hb$. For example, the direct QRD $\Hb = \Qb \Rb$ (see Section \ref{sec.ILSP}) results in the  special case $\yb = \Qb^H\rb$ and $\Rb = \Qb^H \Hb$.

{ { \em Theorem 1:} 
Consider SD with fixed $\rho$ ($0 <\rho <\infty$) and let $\Hb$ and $\rb$ be random (potentially statistically dependent). The corresponding $\layerk$th layer complexity $\NN_{\layerk}$, defined in \eqref{eqn.def.Sk}, satisfies 
\begin{equation}\label{eqn.result.asympt.equal}
	\P[\NN_{\layerk} \geq \L] \dotequal   \P\!\left[ \frac{1}{\text{det}(\Rb_{\layerk}^H \Rb_{\layerk})}  \geq \L \right],
	\quad \L \rightarrow \infty
\end{equation}
if all of the following conditions are met: 
\begin{itemize}
\item {\em Statistics of $\Hb$}: The probability density function (pdf) $f(\Hb)$ of $\Hb$ satisfies the scaling property 
\begin{equation}\label{eqn.result.asympt.equal.cond2}
	f(\Hb) \geq \beta f(a \Hb)
\end{equation}
for all $\Hb \in \Cnum^{\rdim\times \tdim}$ and all  $a \in \Rnum$, $a >1$, with some constant $\beta \in \Rnum$, $\beta >0$. 
\item  {\em Statistics of $\Hb$  and preprocessing}: The covering radius $\rcover(\Rb)$ of $\lattice(\Rb)$ satisfies 
\begin{equation}\label{eqn.result.asympt.equal.cond1}
\P\!\left[ \rcover(\Rb) \geq \L \right]  \dotequal \L^{-\infty}, \quad \L \rightarrow \infty. 
\end{equation}
\item {\em Preprocessing}: 
Let $\text{det}(\Rb_{\layerk}^H \Rb_{\layerk}) = g_{\layerk}(\Hb)$ with $g_{\layerk}: \Cnum^{\rdim\times \tdim} \mapsto \Rnum_{+}$ and $\mu(\Rb) = g_{\mu}(\Hb)$ 
with  $g_{\mu}: \Cnum^{\rdim\times \tdim} \mapsto \Rnum_{+}$. The functions $g_{\layerk}(\Hb)$ and $g_{\mu}(\Hb)$ satisfy, respectively,  the scaling properties
\begin{equation}\label{eqn.result.asympt.equal.cond3}
	g_{\layerk}(b\Hb) = b^{\alpha_{\layerk}} g_{\layerk}(\Hb) 
\end{equation}
and 
\begin{equation}\label{eqn.result.asympt.equal.cond4}
	g_{\mu}(b\Hb) = b^{\alpha} g_{\mu}(\Hb)
\end{equation}
 for all $\Hb \in \Cnum^{\rdim \times \tdim}$ and all $b \in \Rnum$, $b > 0$, with some constants $\alpha_{\layerk}, \alpha \in \Rnum$, $\alpha_{\layerk}>0$, $\alpha>0$. 
\end{itemize}
Proof: See Appendix. $\Box$ 

\subsubsection*{Discussion}
Theorem 1 states that the TB of $\P[\NN_{\layerk} \geq \L]$  is  fully characterized by the TB of 
$\P[1/\text{det}(\Rb_{\layerk}^H \Rb_{\layerk}) \geq \L]$ provided the conditions 
\eqref{eqn.result.asympt.equal.cond2}\,--\,\eqref{eqn.result.asympt.equal.cond4} are satisfied. It is immediate that the TB of $\P[\NN_{\layerk} \geq \L]$ then 
depends neither on the statistics of $\rb$ nor on the particular fixed choice of $\rtwo$. Consequently, 
 $\rb$ and $\rtwo$ 
will influence 
$\P[\NN_{\layerk} \geq \L]$ (for example, a larger $\rtwo$ will certainly result in a larger value of $\P[\NN_{\layerk} \geq \L]$) but do {\em not} affect the corresponding complexity tail exponent. 
The conditions \eqref{eqn.result.asympt.equal.cond2}\,--\,\eqref{eqn.result.asympt.equal.cond4} constitute fairly general 
requirements on the statistics of the 
lattice basis matrix $\Hb$ and on the preprocessing algorithm. For example, for direct QRD or conventional LR (see, e.g., \cite{agrell_it02}),  it can be shown \cite{seethal_IT09} that all the conditions above  are satisfied if the entries of $\Hb$ are jointly Gaussian-distributed with arbitrary non-singular covariance matrix and arbitrary finite mean, i.e., for $\Hb$ being a Rayleigh- or Ricean-fading MIMO channel with (non-singular) covariance matrix. 

It is interesting to note that  $\text{det}(\Rb_{\layerk}^H \Rb_{\layerk})$ is the volume of a fundamental region of $\lattice(\Rb_{\layerk})$  \cite{conway88}. 
A well-known approximation for $\NN_{\layerk}$ is given by \cite{GruberWills93}  
\begin{equation*}\label{eqn.volume.approximation}
\widehat{\NN}_{\layerk} = \frac{V_{\layerk}(\rtwo)}{\text{det}(\Rb_{\layerk}^H \Rb_{\layerk})} 
\end{equation*}
where 
\begin{equation}\label{eqn.volume.sphere}
	V_{\layerk}(\rho) = \frac{\pi^{\layerk} (\rtwo^2)^{\layerk}}{\layerk !}
\end{equation}
is the volume of a hypersphere in $\layerk$ complex-valued dimensions. This approximation simply counts the number of fundamental regions (each occupied by exactly one lattice point) that fit into the $\layerk$-dimensional search sphere and becomes exact if an averaging of  $\NN_{\layerk}$ is performed over $\yb_{\layerk}$ uniformly distributed over a 
fundamental region of $\lattice(\Rb_{\layerk})$ \cite{GruberWills93}. Motivated by this result, $\widehat{\NN}_{\layerk}$ has been used in \cite{agrell_it02} and \cite{banih_98} to assess the complexity of various SD variants. 
 For the TB, it immediately follows that \eqref{eqn.result.asympt.equal} can equivalently be written as $\P[\NN_{\layerk} \geq \L] \dotequal   \P\big[ \widehat{\NN}_{\layerk}  \geq \L \big]$, $\L \rightarrow \infty$, and no averaging argument is required. 

\section{Tail Exponents for i.i.d.\ Gaussian $\Hb$}\label{sec.iid.Gaussian.case}

Specializing Theorem 1 to lattice basis matrices $\Hb$ whose entries are i.i.d.\ $\CN(0,\sigma_{\!H}^2)$ (the model typically used in MIMO detection) leads to particularly interesting results. 
In this case, the pdf of $\Hb$ is given by 
$f(\Hb) = c_{1}\, e^{-c_{2} \|\Hb\|_{\text{F}}^2}$ with some constants $c_{1}, c_{2} >0$, which directly implies that condition \eqref{eqn.result.asympt.equal.cond2} is satisfied with $\beta = 1$.

\subsection{Tail Exponents for Direct QRD}\label{sec.tail.exponents.for.direct.QRD}
For direct QRD of $\Hb$ (i.e., $\Rb$ is obtained through $\Hb = \Qb \Rb$), $\rcover^{2}(\Rb)$ 
can be upper-bounded as (see, e.g., \cite[Prop.\ 1]{banih_98} extended to the complex-valued case) $\rcover^2(\Rb) \leq  \frac{1}{2}\sum_{i=1}^{\tdim} R_{i,i}^2 = z^2$. It  follows from 
\cite[Lemma 2.1]{tulino04} that $z$ is a $\chi$-distributed RV, which, upon noting that $\P\left[ \rcover(\Rb) \geq \L \right] \leq \P\big[  z \geq \L \big]$ implies that condition \eqref{eqn.result.asympt.equal.cond1} is satisfied (see  \cite{seethal_IT09} for a detailed proof of this statement). Condition \eqref{eqn.result.asympt.equal.cond3} 
 is verified by observing that the QRD of $b\Hb$ results in $b\Rb$, which gives 
$g_{\layerk}(b\Hb) = b^{2\layerk} g_{\layerk}(\Hb)$. Condition \eqref{eqn.result.asympt.equal.cond4} is shown to be satisfied by 
noting that \eqref{eqn.covering.radius} implies $\rcover(b\Rb) = b\rcover(\Rb)$ and hence 
$g_{\mu}(b\Hb) = b g_{\mu}(\Hb)$.  Therefore, all the conditions of Theorem 1 are met. Finally, using results from \cite{zheng02}, the TB of 
the distributions of the layer complexities of SD for direct QRD and i.i.d.\ Gaussian $\Hb$ can be established as \cite{seethal_IT09} 
\begin{equation}\label{eqn.dist.complexity.layer.final}
\P[\NN_{\layerk} \geq \L] \dotequal  \L^{-(\rdim-\tdim+1)}, \quad\! \L \rightarrow \infty, \quad  \layerk = 1, \dots, \tdim
\end{equation}
and, consequently, 
\begin{equation}\label{eqn.dist.complexity.total.final}
\P[\NNtotal \geq \L] \dotequal  \L^{-(\rdim-\tdim+1)}, \quad \L \rightarrow \infty. 
\end{equation}
We conclude that the distributions of the layer and total complexities are of Pareto-type with tail exponents $\cexp_{\layerk} = \cexp   = \rdim-\tdim+1$, $\layerk=1,\dots,\tdim$. These results show that increasing $\rdim$ (e.g., the number of receive antennas in a MIMO context) for given $\tdim$ (e.g., the number of transmit antennas) results in improved tail exponents. 

\subsection{Tail Exponents for LR-Based Preprocessing}\label{sec.complexity.tail.exponents.for.LR-SD}

We define LR-based preprocessing (see, e.g., \cite{agrell_it02}) as applying, prior to the QRD, the transformation 
$\Hbr = \Hb \Tb$, where $\Tb$ is an  $\tdim \times \tdim$ unimodular matrix, i.e., $T_{i,j} \in \CZnum$, $\forall i,j$,  and $|\text{det}(\Tb)| = 1$. 
The matrix $\Tb$ is obtained, for example, through the LLL algorithm \cite{lenstra82}, which finds a basis matrix $\Hbr$ of the lattice $\lattice(\Hb)$ that is ``closer'' than $\Hb$ to an orthogonal matrix. Another important preprocessing technique is LS (e.g., with the V-BLAST algorithm \cite{golden99}), which is just a special case of LR obtained by restricting $\Tb$ to be a permutation matrix.

The triangularized form of the CLP problem based on LR preprocessing is given by \eqref{eqn.ILSP.QR} with 
$\Rb$ and $\yb$ replaced by $\Rbr$ and $\ybr = \Qbr^H \rb$, respectively, where $\Qbr$ and $\Rbr$  are the QR-factors of $\widetilde{\Hb}$, i.e., $\widetilde{\Hb} =  \Qbr \Rbr$. If we denote the corresponding solution of \eqref{eqn.ILSP.QR} as $\widetilde{\db}$, the final solution of \eqref{eqn.ILSP} is 
$\widehat{\db} = \Tb \widetilde{\db}$. We now consider Theorem 1 with $\Rb$ replaced by $\Rbr$. 
Let us write $\Rb \Tb  = \QbrR\, \RbrR$, where $\QbrR$ and $ \RbrR$ are the QR-factors of $\Rb \Tb$.
Noting that $\widetilde{\Hb} = \Qb \Rb \Tb = \Qbr \Rbr$, we obtain $\Qb\QbrR\, \RbrR =  \Qbr \Rbr$. Since $\Qb\QbrR$ is unitary and 
the QR-factors are unique, it follows that $\Qb\QbrR = \Qbr$ and $ \RbrR =  \Rbr$, which implies 
\begin{equation}\label{eqn.Rd.Rlr}
	\Rb  = \QbrR\, \Rbr \Tb^{-1}. 
\end{equation}
As $\Tb^{-1}$ is unimodular (since $\Tb$ is unimodular) and $\QbrR$ is unitary,  
it can be verified that 
$\rcover(\Rb) = \rcover(\Rbr)$ and $\text{det}(\Rbr^H \Rbr) = \text{det}(\Rb^H \Rb)$. 
 Due to $\rcover(\Rb) = \rcover(\Rbr)$ and the results for direct QRD in Section \ref{sec.tail.exponents.for.direct.QRD}, 
conditions \eqref{eqn.result.asympt.equal.cond1} and  \eqref{eqn.result.asympt.equal.cond4} are satisfied for LR-based preprocessing. 

\subsubsection*{LR for $\layerk = \tdim$}   
Due to $\text{det}(\Rbr^H \Rbr) = \text{det}(\Rb^H \Rb)$, condition \eqref{eqn.result.asympt.equal.cond3} is satisfied for $\layerk = \tdim$ and LR-based preprocessing. 
Now, applying Theorem 1, we can immediately conclude that LR-based processing results in the {\em same} $\tdim$th layer complexity tail exponent as direct QRD,  i.e., 
\begin{equation}\label{eqn.Mth.layer.complexity}
\P[\NN_{\tdim} \geq \L] \dotequal  \L^{-(\rdim-\tdim+1)}, \quad \L \rightarrow \infty
\end{equation}
or, equivalently,  $\cexp_{\tdim} = \rdim-\tdim+1$. From  \eqref{eqn.cexp.total.cexpk}, we can therefore conclude that $\cexp \leq \rdim-\tdim+1$ for LR-based preprocessing, which shows that LR (including LS)  
 does {\em not} improve (i.e., increase) the total complexity tail exponent as compared to that obtained for direct QRD. 
 
\subsubsection*{LR for $\layerk < \tdim$}
It can be shown \cite{seethal_IT09} that all LR algorithms delivering a unimodular transformation matrix $\Tb$, which is invariant to 
a positive scaling of $\Hb$, i.e., $\Hb$ and $b\Hb$ for all $b\in \Rnum$, $b>0$, result in the same $\Tb$, satisfy condition \eqref{eqn.result.asympt.equal.cond3} for $\layerk = 1,\dots,\tdim$.  
We note that this is the case for any reasonable LR algorithm we are aware of. Prominent examples are the LLL algorithm \cite{lenstra82} and LS according to the V-BLAST algorithm  \cite{golden99}. 
Here, 
 Theorem 1 therefore implies  that 
 \begin{equation}\label{eqn.result.asympt.equal.LR}
	\P[\NN_{\layerk} \geq \L] \dotequal  \P\bigg[ \frac{1}{\text{det}(\Rbr_{\layerk}^H \Rbr_{\layerk})}  \geq \L \bigg], \quad \L \rightarrow \infty
\end{equation} 
for $\layerk = 1,\dots,\tdim$.  

\subsubsection*{LR Based on LLL}\label{sec.lr.based.on.lll}
For LR carried out through the LLL algorithm  \cite{lenstra82} (see \cite{Mow05} for  its complex-valued extension), based on \eqref{eqn.result.asympt.equal.LR}, one can show the more specific result 
 \cite{seethal_IT09}
\begin{equation*}
\P[\NN_{\layerk} \geq \L] \dotleq \L^{-\frac{\rdim}{\layerk}},  \quad \L \rightarrow \infty, \quad \layerk = 1,\dots, \tdim
\end{equation*}
or, equivalently, $\cexp_{\layerk} \geq \rdim/\layerk$. Compared  with $\cexp_{\layerk} = \rdim - \tdim+1$ for direct QRD (cf.\ \eqref{eqn.dist.complexity.layer.final}), we can conclude that 
 LLL preprocessing improves (i.e., increases) the tail exponents up to layer $\layerk \leq \lceil \rdim/(\rdim-\tdim+1)\rceil-1$.  
 In the following, consider $\rdim = \tdim$. 
We have $\cexp_{\layerk} = \tdim/\layerk > 1$, $\layerk = 1, \dots, \tdim-1$, and 
 $\cexp_{\tdim} = 1$ (see \eqref{eqn.Mth.layer.complexity}), which, in this case, establishes that the 
 TB of the distribution of the total complexity of SD with LLL preprocessing is dominated by the TB of the distribution of the $\tdim$th layer complexity; 
 in particular, we have 
 $\cexp = \cexp_{\tdim} = 1$, as in the case of direct QRD (cf.\ \eqref{eqn.dist.complexity.total.final}).  
 
\section{Numerical Results}\label{sec.numerical.results.MIMO}
We consider SD for data detection in $\rdim \times \tdim$ MIMO wireless systems, where  
 $   \rb \,=\, \Hb\db' + \wb$ with the entries of $\Hb$ and $\wb$ assumed i.i.d.\ 
$\CN(0,1/\tdim)$ and i.i.d.\  $\CN(0,\wvar)$, respectively, and with the transmitted vector $\db' \in (\CZnum)^{\tdim}$. 
We choose the radius $\rtwo$ in \eqref{eqn.SC.qr} such that $\db'$ is found by the SD algorithm 
with probability  $0.99$ and for $1/\wvar$ we assume a value of $15\,$dB.  We note that the complexity of SD is random in $\Hb$ and $\wb$ and does not depend on $\db'$. 

\begin{figure}[t]
\begin{center}
\unitlength1mm
\begin{picture}(85,45)
\put(8,-3){\resizebox{7.1cm}{!}{\input N_k_45x4_15dB_001_comp.pdftex_t}}
\end{picture}
\caption{Distribution of total complexity $\P[\NN \geq \L]$ of SD with direct QRD, V-BLAST LS, and LLL preprocessing for $4\times 4$ and $5\times 4$ MIMO systems.}
\vspace{-0.5cm}
\label{fig.SD.comp}
\end{center}
\end{figure}
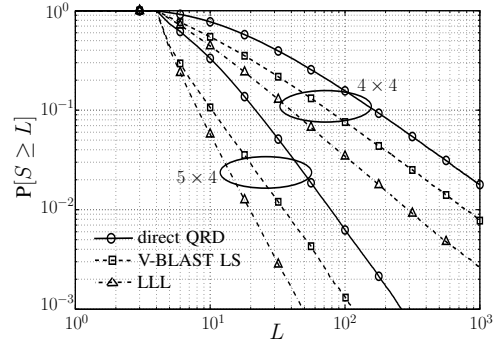

 Fig.\ \ref{fig.SD.comp} shows the distribution of the total complexity $\P[\NN \geq \L]$ in double log-scale for SD with direct QRD, V-BLAST LS  \cite{golden99}, and with (complex-valued) LLL preprocessing 
 \cite[with parameter 
 $\delta = 3/4$]{Mow05} for $4\times 4$
  and 
  $5\times 4$ 
   MIMO systems.
 We can see that the results reflect our analytic findings. For example, for direct QRD in the $4\times 4$ case, the distribution of the total complexity in Fig.\ \ref{fig.SD.comp}  exhibits a large-$\L$ behavior of $\L^{-1}$ as predicted by  \eqref{eqn.dist.complexity.total.final} for $\rdim = \tdim$. Furthermore,  it can be seen that adding one receive antenna, indeed, improves the TB and leads to a large-$\L$ behavior of $\L^{-2}$ (cf.\ \eqref{eqn.dist.complexity.total.final}). Finally, the numerical results indicate that LLL preprocessing and V-BLAST LS do reduce the complexity, as compared to direct QRD, but do not improve the tail exponents. 
 
\section*{Acknowledgment}
The authors would like to thank C.\ Ak\c{c}aba and P.\ Coronel for helpful discussions.

\appendices 
\section*{Appendix\\ Proof of Theorem 1}
\label{app.general.result}
The proof of Theorem 1 is based on separately establishing the exponential lower bound $\P[\NN_{\layerk} \geq \L]  \dotgeq  \P\!\left[ 1/\text{det}(\Rb_{\layerk}^H \Rb_{\layerk}) \geq  \L  \right]$, 
$\L \rightarrow \infty$, and the exponential upper bound $\P[\NN_{\layerk} \geq \L]  \dotleq  \P\!\left[ 1/\text{det}(\Rb_{\layerk}^H \Rb_{\layerk}) \geq  \L  \right]$, $\L \rightarrow \infty$, which then combine to   
$\P[\NN_{\layerk} \geq \L]  \dotequal \P\!\left[ 1/\text{det}(\Rb_{\layerk}^H \Rb_{\layerk}) \geq  \L  \right]$. 

\subsection{Exponential Lower Bound}
We start by noting that  \cite[Ch.\ 3.2, Eq.\ (3.3)]{GruberWills93}
\[
 \NN_{\layerk} \geq \frac{V_{\layerk}(\rtwo) - \rcover(\Rb_{\layerk}) A_{\layerk}(\rtwo)}{\text{det}(\Rb_{\layerk}^H \Rb_{\layerk})}
\]
where $V_{\layerk}(\rtwo)$ and $A_{\layerk}(\rtwo)$ denote the volume (cf.\ \eqref{eqn.volume.sphere})
 and the surface area of the search sphere at layer $\layerk$, respectively.\footnote{Note that condition \eqref{eqn.result.asympt.equal.cond1} implies full-rank $\Rb_{\layerk}$ with probability one. In particular, 
 $\text{det}(\Rb_{\layerk}^H \Rb_{\layerk}) > 0$  and $\rcover(\Rb_{\layerk}) < \infty$ with probability one.
  However, it is straightforward to show that Theorem 1 also holds in the case where $\Rb_{\layerk}$ is rank-deficient with non-zero probability, which leads to $\P[\NN_{\layerk} \geq \L]  \dotequal \P\!\left[ 1/\text{det}(\Rb_{\layerk}^H \Rb_{\layerk}) \geq  \L  \right] \dotequal  \L^0$, $\L \rightarrow \infty$.} Using $\rcover(\Rb_{\layerk}) \leq \rcover(\Rb)$, $\layerk = 1,\dots, \tdim$, \cite{seethal_IT09}, we obtain  
\[
\P[\NN_{\layerk} \geq \L]  \geq \P\left[ \frac{V_{\layerk}(\rtwo) - \rcover(\Rb)  A_{\layerk}(\rtwo)}{\text{det}(\Rb_{\layerk}^H \Rb_{\layerk})}  \geq \L \right]. 
\]
Consider a constant $c \in \Rnum$, $c >0$, such that $V_{\layerk}(\rtwo) - c A_{\layerk}(\rtwo) > 0$ and define $c' = V_{\layerk}(\rtwo) - c A_{\layerk}(\rtwo) >0$. We then have 
\begin{equation}\label{eqn.lower.bound.zw2}
   \P[\NN_{\layerk} \geq \L] \geq \P\left[ \Hb \in {\cal B} \right ]
\end{equation}
where
$ 
{\cal B} = \left\{\Hb \!: \left(\frac{c'}{g_{\layerk}(\Hb) }  \geq \L\right) \!\cap (g_{\mu}(\Hb) \leq c) \right\}
$  
with $\text{det}(\Rb_{\layerk}^H \Rb_{\layerk}) = g_{\layerk}(\Hb)$ and $\rcover(\Rb) = g_{\mu}(\Hb)$.  
With property \eqref{eqn.result.asympt.equal.cond2}, we further obtain 
\begin{align*}
\P\!\left[ \Hb \in {\cal B} \right ] = \int_{\Hb \in {\cal B}} \!\!\!f(\Hb) d\Hb \geq \beta \int_{\Hb \in {\cal B}} \!\!\! f(\L^{\delta}\Hb) d\Hb
\end{align*}
for all $\delta>0$, $\L>1$, and some $\beta >0$.  Performing the change of variables $\Hb' = \L^{\delta}\Hb$ and invoking conditions \eqref{eqn.result.asympt.equal.cond3} and \eqref{eqn.result.asympt.equal.cond4} yields  
 \begin{equation}\label{eqn.lower.bound.zw3}
\P\!\left[ \Hb \in {\cal B} \right ] \geq \beta\, \L^{-2\tdim\rdim\delta} \,\P\!\left[ \Hb \in {\cal B'}\right ]
\end{equation}
where 
\[
{\cal B'} = \left\{ \Hb \!: \!\left(\frac{c'}{g_{\layerk}(\Hb)}  \geq  \L^{1-\delta\alpha_{\layerk}} \right) \cap (g_{\mu}(\Hb) \leq c \L^{\delta\alpha}) \right\}.
\]
Next, noting that  for two events $A_{1}$ and $A_{2}$,  by the inclusion-exclusion principle, $\P[A_{1} \cap A_{2}]\geq \P[A_{1}]-\P[\bar{A}_{2}]$, where $\bar{A}_{2}$ denotes the complementary event of $A_{2}$, we get
\[
 \P\!\left[ \Hb \in {\cal B}' \right ] \geq  \P\!\left[  \frac{c'}{g_{\layerk}(\Hb)}  \geq  \L^{1-\delta\alpha_{\layerk}}  \right] -\P\!\left[g_{\mu}(\Hb)   > c \L^{\delta\alpha} \right].
\]
Now \eqref{eqn.result.asympt.equal.cond1} with $\mu(\Rb) = g_{\mu}(\Hb)$ and $\delta, \alpha > 0$ implies 
$\P\!\left[g_{\mu}(\Hb) > c \L^{\delta\alpha} \right] \dotequal\L^{-\infty}$, $\L \rightarrow \infty$, 
which, together with \eqref{eqn.lower.bound.zw2} and \eqref{eqn.lower.bound.zw3},  yields
 \begin{equation*} 
  \P[\NN_{\layerk} \geq \L]  \dotgeq  \L^{-2\tdim\rdim\delta}\,  \P\!\left[    \frac{c'}{g_{\layerk}(\Hb)}  \geq  \L^{1-\delta\alpha_{\layerk}}  \right], \quad \L \rightarrow \infty.
 \end{equation*}
Let us write $\P\!\left[1/g_{\layerk}(\Hb)  \geq  \L  \right] \dotequal \L^{-a}$, $\L \rightarrow \infty$, for some constant $a\geq0$. We then have  
 $ \P[\NN_{\layerk} \geq \L]  \dotgeq   \L^{-2\tdim\rdim\delta-(1-\delta\alpha_{\layerk})a}$, $\L \rightarrow \infty$.  As this result holds for arbitrarily small values of $\delta$, 
 we can conclude that $ \P[\NN_{\layerk} \geq \L]  \dotgeq \L^{-a} \dotequal  \P\!\left[1/g_{\layerk}(\Hb)  \geq  \L  \right]$, $\L \rightarrow \infty$, which establishes the exponential lower bound.

\subsection{Exponential Upper Bound}
From \cite[Ch.\ 3.2, Eq.\ (3.3)]{GruberWills93} 
\[
 \NN_{\layerk} \leq \frac{V_{\layerk}(\rtwo+\rcover(\Rb_{\layerk}))}{\text{det}(\Rb_{\layerk}^H \Rb_{\layerk})}
\]
which, again using $\mu(\Rb_{\layerk}) \leq \mu(\Rb)$, $\layerk = 1,\dots, \tdim$, results in  
\begin{equation}\label{eqn.upper.bound.start}
\P[\NN_{\layerk} \geq \L]  \leq \P\left[ \frac{V_{\layerk}(\rtwo+\rcover(\Rb))}{\text{det}(\Rb_{\layerk}^H \Rb_{\layerk})}  \geq \L \right]. 
\end{equation}
\sloppy Note that $\P[x y  \geq \L]   =  
	\P\!\left[ (x y \geq \L) \cap  (y < \L^\delta) \right]  + \P\!\left[(x y \geq \L) \cap (y \geq \L^\delta)\right] 
	 \leq  \P\!\left[x \geq \L^{1-\delta} \,\right] + \P\!\left[y \geq \L^\delta\,\right]$
for any two RVs $x, y \in \Rnum$ and any constant  $\delta \in \Rnum$, $0<\delta<1$. Applying this to \eqref{eqn.upper.bound.start} with $x = 1/\text{det}(\Rb_{\layerk}^H \Rb_{\layerk})$ and $y = V_{\layerk}(\rtwo+\rcover(\Rb))$, we get  
\begin{align} 
		\P[\NN_{\layerk} \geq \L]  & \leq   \P\!\left[ \frac{1}{\text{det}(\Rb_{\layerk}^H \Rb_{\layerk})}  \geq \L^{1-\delta} \,\right] \nonumber  \\[2mm]
		& \hspace{0.8cm}+ \P\!\left[V_{\layerk}(\rtwo+\rcover(\Rb)) \geq \L^\delta\,\right]. \nonumber 
\end{align}
With \eqref{eqn.volume.sphere} and the binomial theorem, we can write 
\[
V_{\layerk}\big(\rtwo+\rcover(\Rb)\big) =  \frac{\pi^{\layerk}}{\layerk !} \sum_{i = 0}^{2\layerk} {2\layerk \choose i} \rtwo^{2\layerk-i}  \big(\rcover(\Rb)\big)^{\!i}
\]
which, using  $\P\!\left[\sum_{i=1}^{M} x_{i} \geq \L\right] \leq\sum_{i=1}^{M} \P[ x_{i} \geq L/M]$ for any set of RVs $\{x_{i}\}_{i=1}^{\tdim}$ yields \vspace{-2mm}
\begin{equation*}\label{eqn.upper.bound.trace.covering}
\P\!\left[V_{\layerk}(\rtwo+\rcover(\Rb))  \geq \L^{\delta} \right]  \dotleq  \sum_{i = 0}^{2\layerk} \P\!\left[ \big(\rcover(\Rb)\big)^{\!i}  \geq \L^{\delta} \right], \quad \L \rightarrow \infty. 
\end{equation*} 
Property \eqref{eqn.result.asympt.equal.cond1} (for the terms corresponding to $i>0$) and $\P[c'' \geq L ] \leq e^{-(\L-c'')} \dotequal \L^{-\infty}$, $\L \rightarrow \infty$, for any constant $c''\geq0$ (for the term corresponding to $i=0$) now directly imply 
$\P\!\left[ V_{\layerk}(\rtwo+\rcover(\Rb))  \geq \L^{\delta} \right]  \dotequal \L^{-\infty}$, $\L \rightarrow \infty$, 
and, hence, 
\[
\P[\NN_{\layerk} \geq \L]   \dotleq   \P\!\left[ \frac{1}{\text{det}(\Rb_{\layerk}^H \Rb_{\layerk})}  \geq \L^{1-\delta} \,\right], \quad \L \rightarrow \infty. 
\]
As before, writing $\P\!\left[  1/\text{det}(\Rb_{\layerk}^H \Rb_{\layerk})  \geq  \L  \right] \dotequal \L^{-a}$, $\L \rightarrow \infty$, for some constant $a\geq0$, we get 
$\P[\NN_{\layerk} \geq \L]    \dotleq  \L^{-(1-\delta)a}$, $\L \rightarrow \infty$. 
As this result holds for arbitrarily small values of $\delta$, we can conclude that 
$\P[\NN_{\layerk} \geq \L]  \dotleq  \L^{-a} \dotequal \P\!\left[   1/\text{det}(\Rb_{\layerk}^H \Rb_{\layerk})  \geq  \L  \right]$,  $\L \rightarrow \infty$, 
which establishes  the exponential upper bound. 

\renewcommand{\baselinestretch}{.955}\footnotesize
\bibliographystyle{IEEEtran}      %
\bibliography{/Users/Dom/Documents/work/tex/tf-zentral_loc,/Users/Dom/Documents/work/tex/bib_see}
 
\end{document}